# INSPIRE: The entry point to Europe's Big Geospatial Data Infrastructure

Marco Minghini[1], Vlado Cetl[1], Alexander Kotsev[1], Robert Tomas[1], Michael Lutz[1]

[1] European Commission, Joint Research Centre (JRC), 21027 Ispra, Italy

marco.minghini@ec.europa.eu, vlado.cetl@ec.europa.eu, alexander.kotsev@ec.europa.eu, robert.tomas@ec.europa.eu, michael.lutz@ec.europa.eu

*The views expressed are purely those of the authors and may not in any circumstances be regarded as stating an official position of the European Commission.*

## ABSTRACT

Initiated in 2007, the INSPIRE Directive has set a legal framework to create a European-wide Spatial Data Infrastructure (SDI) to support the European Union (EU) environmental policies. This chapter analyses the INSPIRE infrastructure from a Big Geospatial Data perspective, describing how data is shared in an interoperable way by public sector organisations in the EU Member States and how it is made available in and accessible within the infrastructure. The INSPIRE Geoportal, which is the entry point to the whole infrastructure, is presented in detail. To justify its nature of a Big Geospatial Data infrastructure, the characteristics of INSPIRE data are mapped to those of Big Data's six 'Vs'. Despite many good results achieved in terms of data sharing, some challenges still remain related to data consumption from the user side. The chapter concludes with a dedicated discussion on how INSPIRE, and traditional SDIs in general, should evolve into modern data ecosystems to address these challenges while also embracing the modern practices of data sharing through the web.

# 1. INTRODUCTION

Technological advancements in the last decade have enabled governments, businesses and citizens to produce and collect increasingly larger amounts of data. The availability of personal digital devices with built-in sensors increased while their price significantly decreased, bringing the chance to collect multitudes of data in a simple and fast way to everyone's reach. This unprecedented large and heterogeneous amount of data collected at exceptional scales and speeds has subsequently led to the establishment of the term Big Data, which has currently become ubiquitous in many areas. Multiple definitions of Big Data are available, which bring together different concepts such as volume, variety, cloud, technology, storage, analytics, processing, information, and transformation. According to the common formal definition proposed by De Mauro et al. (2015) Big Data represents the "Information assets characterised by such a High Volume, Velocity and Variety to require specific Technology and Analytical Methods for its transformation into Value". The exploitation of Big Data is often connected to cloud computing platforms, which are composed of data infrastructures put in place in order to store and manage data, high-bandwidth networks to transport data, and high-performance computers to process data (European Commission 2016).

It is clear that the data revolution that is underway is already reshaping how knowledge is produced, business conducted and governance enacted (Kitchin 2014). The main components of this data revolution are digitalisation, big data, open data and data infrastructures. These components, whose impact is already visible in science, business, government and civil society are addressed by new and emerging fields, such as data science, social computing, and artificial intelligence.

In the geospatial arena, Information and Communication Technology (ICT) developments have been continuously adopted as well. Geospatial data management started with the development of Geographic Information Systems (GIS) which in turn evolved into Spatial Data Infrastructures (SDIs), and consequently in Spatial Knowledge Infrastructures (SKIs). SDIs developments started on the one hand with legally binding governmental initiatives and on the other hand with more business-oriented initiatives driven by the private sector. Typical examples of the former are National Spatial Data Infrastructures (NSDIs) such as the one in the US (Clinton 1994) and European Spatial Data Infrastructures (ESDI), which in the European Union (EU) are driven by the INSPIRE[1] (Infrastructure for Spatial Information in Europe) Directive (European Parliament and Council, 2007). In addition to the EU Member States (MS) and European Free Trade Association (EFTA) countries, candidate and potential candidate countries (e.g. Western Balkans) and some European neighbourhood countries (e.g. Ukraine and Moldova) are also building their NSDIs in accordance with INSPIRE (Cetl et al. 2014). The latter group of SDIs is represented by the mapping frameworks from commercial surveying companies including Google Maps, Microsoft Bing Maps, and HERE Maps. There is also a third group of GDIs driven by crowdsourced initiatives such as OpenStreetMap[2], which has built the largest, most diverse and most detailed open geospatial database to date (Mooney and Minghini, 2017) and whose quality can equal that of authoritative data (see e.g. Haklay 2010, Girres and Touya 2010, Fan et al. 2014, Brovelli et al. 2016). There is no doubt that all these initiatives have triggered the creation of Big Geospatial Data. There is also no doubt that all of them are interrelated and their user bases are becoming more and more similar (see e.g. Köbben and Graham 2009, Minghini et al. 2019). Those heterogeneous initiatives combined, have huge potential for creating synergies in Big Geospatial Data management.

---

[1] https://inspire.ec.europa.eu
[2] https://www.openstreetmap.org

In this chapter our emphasis is on governmental, i.e. authoritative geospatial data in the EU. These are officially recognised, quality-certified data provided by authoritative sources such as Environmental Protection Authorities (EPAs) and National Mapping and Cadastral Agencies (NMCAs). Authoritative geospatial data in the EU are managed and shared through NSDIs, interlinked in a European Big Geospatial Data infrastructure that is to a large extent shaped by the legal provisions of the INSPIRE Directive. Structurally, the chapter is organised as follows. Section 2 offers an overview of the current Big Data initiatives started by the European Commission (EC), with special focus on those characterised by a geospatial component and their relation with INSPIRE. A more detailed introduction to the INSPIRE legal, technical and organisational framework as well as the state of play of INSPIRE is provided in Section 3. Particular attention is placed on the INSPIRE Geoportal, which is the entry point of the whole infrastructure. This is followed by Section 4, which maps the main characteristics of Big Data, expressed by the popular six Vs, to the main features of the INSPIRE infrastructure, thus proving its nature of a Big (Geospatial) Data infrastructure. A number of open issues in making full use of the INSPIRE infrastructure from a user perspective are then listed. Finally, Section 5 concludes the chapter by reflecting on those issues and the lessons learnt from the INSPIRE implementation from a Big Data perspective, and outlining some potential evolutions from a traditional SDI to a modern data ecosystem.

## 2. BIG DATA IN THE EU

According to the Digital Single Market strategy of the European Commission[3], data represents a key asset for the economy and society similar to the traditional categories of human and financial resources. The need to make sense of Big Data – regardless of their nature (geospatial, statistics, weather, research, transport, energy, or health) and source (public services, connected objects, private sector, citizens, research) – is leading to innovations in technology, development of new tools and new skills.

At the EU level, the response to the potential of using the cloud as a platform for Big Data exploitation resides in the European Cloud Initiative (European Commission 2016), which aims to interconnect the existing EU data infrastructures and coordinate their support, ensuring that the sharing of data and the capacity to exploit them are maximised. The European Cloud Initiative is based on the Digital Single Market strategy and several other EU initiatives addressing Big Data, including the 2012 European Cloud Strategy (European Commission 2012a), the High Performance Computing (HPC) Strategy (European Commission 2012b), and the policy developed in the Communication on Big Data (European Commission 2014). A number of reasons are listed why Europe has not yet exploited the full potential of data: non-openness of publicly-funded research data, lack of data interoperability, data fragmentation (i.e. infrastructures scattered across countries and domains), and lack of a European world-class HPC infrastructure. Regarding the lack of data interoperability, the geospatial domain is explicitly mentioned as an exception thanks to the INSPIRE Directive.

The tool envisioned to give Europe a global lead in scientific data infrastructure is the European Open Science Cloud (EOSC; Koski et al. 2015), an open, interoperable, distributed, service-oriented, publicly funded and publicly governed platform connecting networks, data, computing systems, software, tools and services of EU MS to enable gathering, management, analysis, sharing and discovery of scientific data according to the principles of Open Science (European Commission 2015) to ultimately lead to economic and societal innovation. The EOSC will be research-centric, though not specific to any discipline or field, and will aim to make all scientific data produced by the Horizon 2020 Programme open by default according to the principles of Findability, Accessibility, Interoperability and Reusability (FAIR; Wilkinson et al. 2016). However, turning the FAIR principles into reality would require an effort at the level of both the technological infrastructure and the research culture (European Commission Expert Group on FAIR Data 2018). Finally, the European Cloud Initiative already foresees quantum computing as the next breakthrough in supercomputing and secure networking and anticipates the need for Europe to make significant investments to be at the forefront[4].

---

[3] https://ec.europa.eu/digital-single-market/en/big-data

Looking more into the geospatial dimension of Big Data in the EU, in addition to the INSPIRE framework which is separately described in Section 3 there are several other initiatives worth to be mentioned. Copernicus is the EU's Earth Observation (EO) Programme, looking at our planet and its environment for the ultimate benefit of all European citizens[5]. It offers service providers, public authorities and other international organisations a number of Services – focused on atmosphere, marine, land, climate change, security and emergency – based on satellite EO and in situ (non-space) data. The processed data and the information disseminated, both freely and openly accessible, put Copernicus at the forefront of the Geospatial Big Data paradigm. More than 5 million products have been published in the Sentinel repository managed by the European Space Agency (ESA) and more than 100.000 users have downloaded more than 50 PB of data since the system became operational, with 1 PB of data corresponding to about 750.000 datasets (Koubarakis et al. 2019). This volume, as well as the velocity at which data are collected and processed, will increase in the future with the launch of new Sentinel satellites. Copernicus is closely connected with INSPIRE since Copernicus Services need access to openly available, up-to-date and harmonised geospatial information across Europe for production and validation purposes. In turn, many geospatial datasets and services produced by Copernicus are exposed according to the INSPIRE guidelines to maximize their interoperability. In the field of positioning and navigation, a partnership between the European Space Agency (ESA) and the EC has resulted into Galileo, the European Global Navigation Satellite System (GNSS) offering high-quality positioning, navigation and timing services to users across the world[6]. Galileo is fully compatible with the American GPS and Russian GLONASS, thus offering enhanced combined performance; in contrast to them, it is specifically designed to remain under civilian control. Galileo's full operational constellation – still under construction – will consist of 24 operational satellites plus six spares circling Earth in three circular medium-Earth orbits, at an altitude of about 23000 km. Galileo builds upon the success of the European Geostationary Navigation Overlay Service (EGNOS)[7], operational since 2009 to provide safety of life navigation services to aviation, maritime and land-based users over most of Europe. Finally, the EC is participating in the Group on Earth Observations (GEO)[8], a global network of governmental institutions, research organisations, data providers, scientists and experts working together to build a Global Earth Observation System of Systems (GEOSS)[9] aimed at strengthening the monitoring of the Earth and improving decision making. Thanks to the knowledge acquired through Copernicus, Galileo, EGNOS and other programmes, Europe has positioned itself as a global force in the field of EO and in 2019 the European portion of GEO was renamed EuroGEO, and EuroGEOSS was established as the European component of GEOSS, yet again with a clear link to INSPIRE (European Commission 2017).

---

[4] https://ec.europa.eu/digital-single-market/en/news/quantum-technologies-opportunities-european-industry-report-round-table-discussion-and
[5] https://www.copernicus.eu
[6] https://www.gsa.europa.eu/european-gnss/galileo/galileo-european-global-satellite-based-navigation-system
[7] https://www.gsa.europa.eu/egnos/what-egnos
[8] http://www.earthobservations.org/index.php
[9] https://www.earthobservations.org/geoss.php

In the context of a general strategy to build a European data economy[10], and within a framework for digital trust granted by the General Data Protection Regulation (European Parliament and Council 2016), a recent initiative dedicated to the establishment of a common European data space is addressing challenges pertaining to data value chains in the era of Big Data. The related EC Communication defines a data space as "a seamless digital area with the scale that will enable the development of new products and services based on data" and positions public data at the centre of data-driven innovation (European Commission 2018). Additional emphasis is put on ensuring access to publicly-funded data held by private companies, and different data-flows (business-to-business, business-to-government, etc.) that are beneficial for all actors involved. This is further addressed by the recent Open Data Directive (European Parliament and Council 2019), which encourages the FAIR management of EU public and publicly funded data and recognises INSPIRE as a good practice. The Directive also introduces the concept of *high-value datasets*, i.e. datasets with "the potential to (i) generate significant socio-economic or environmental benefits and innovative services, (ii) benefit a high number of users, in particular SMEs, (iii) assist in generating revenues, and (iv) be combined with other datasets"; hence, it requires that such datasets are made available free of charge, in machine-readable formats and provided via Application Programming Interfaces (APIs) and as a bulk download where relevant. The Directive does not provide a full list of such datasets – which is left for future work – but only defines categories of datasets, one of which is the geospatial one.

## 3. INSPIRE State of Play

### 3.1 Legal, technical and organisational framework

The legal framework for INSPIRE has been set by the Directive 2007/2/EC (European Parliament and Council 2007) and related interdependent legal acts, which are called Implementing Rules, in the form of Commission Regulations and Decisions. By design, the INSPIRE infrastructure is built upon the NSDIs established and operated by the EU MS and EFTA countries that are then made compliant with the Implementing Rules, covering its core components: metadata, network services, interoperability of spatial datasets and services, data sharing and monitoring and reporting (Tomas et al. 2015; Cetl et al. 2019). The Implementing Rules for metadata, the interoperability of data themes, the network services (that help to share the infrastructure's content online) and the data sharing are complemented by non-legally binding Technical Guidance documents. These guidelines explain a possible technical approach to fulfill the legal requirements and embed additional recommendations that may help data providers in their implementation for a range of use cases.

The thematic scope of INSPIRE includes 34 cross-sectoral categories, named data themes (see Figure 1), listed in the three annexes of the Directive and reflecting two main types of data: baseline geospatial data (presented in Annex I and partly in Annex II), which define a location reference that the remaining data themes (in Annex III and partly in Annex II) can then refer to.

---

[10] https://ec.europa.eu/digital-single-market/en/policies/building-european-data-economy

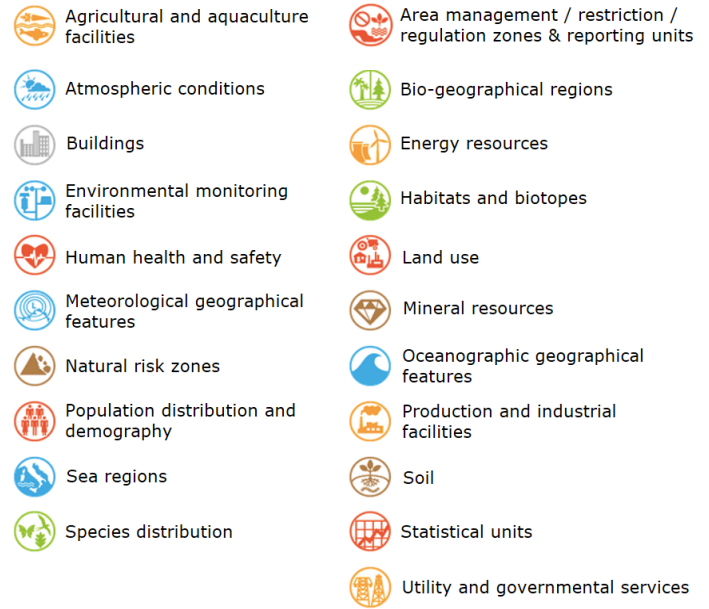

Figure 1. INSPIRE themes, organised in three Annexes. Source: European Commission, Joint Research Centre.

Data and metadata are shared through web-based services, referred to as network services (European Commission 2009), based on a Service Oriented Architecture (SOA)[11] approach (see Figure 2). Network services are implemented through well-established international standards for geospatial interoperability, mainly developed by the Open Geospatial Consortium (OGC)[12]. Technical Guidance documents illustrate how data providers can establish access to metadata for Discovery Services through the Catalogue Service for the Web (CSW)[13]. Similarly for View Services, the interactive visualisation of georeferenced content involves guidelines using the Web Map Service (WMS)[14] and Web Map Tile Service (WMTS)[15] standards. Download Services also have guidelines that recommend the use of Atom feeds[16], Web Feature Service (WFS)[17], Web Coverage Service (WCS)[18] and Sensor Observation Service (SOS)[19], for appropriate types of data. There are also various Transformation Services defined, which can support coordinate and data transformations. In addition to all the above, there are generic services (registry and other spatial data services), that are implemented on a national as well as European level.

---

[11] https://www.opengroup.org/soa/source-book/soa/p1.htm
[12] http://www.opengeospatial.org/
[13] https://www.opengeospatial.org/standards/cat
[14] https://www.opengeospatial.org/standards/wms
[15] https://www.opengeospatial.org/standards/wmts
[16] https://validator.w3.org/feed/docs/atom.html
[17] https://www.opengeospatial.org/standards/wfs
[18] https://www.opengeospatial.org/standards/wcs
[19] https://www.opengeospatial.org/standards/sos

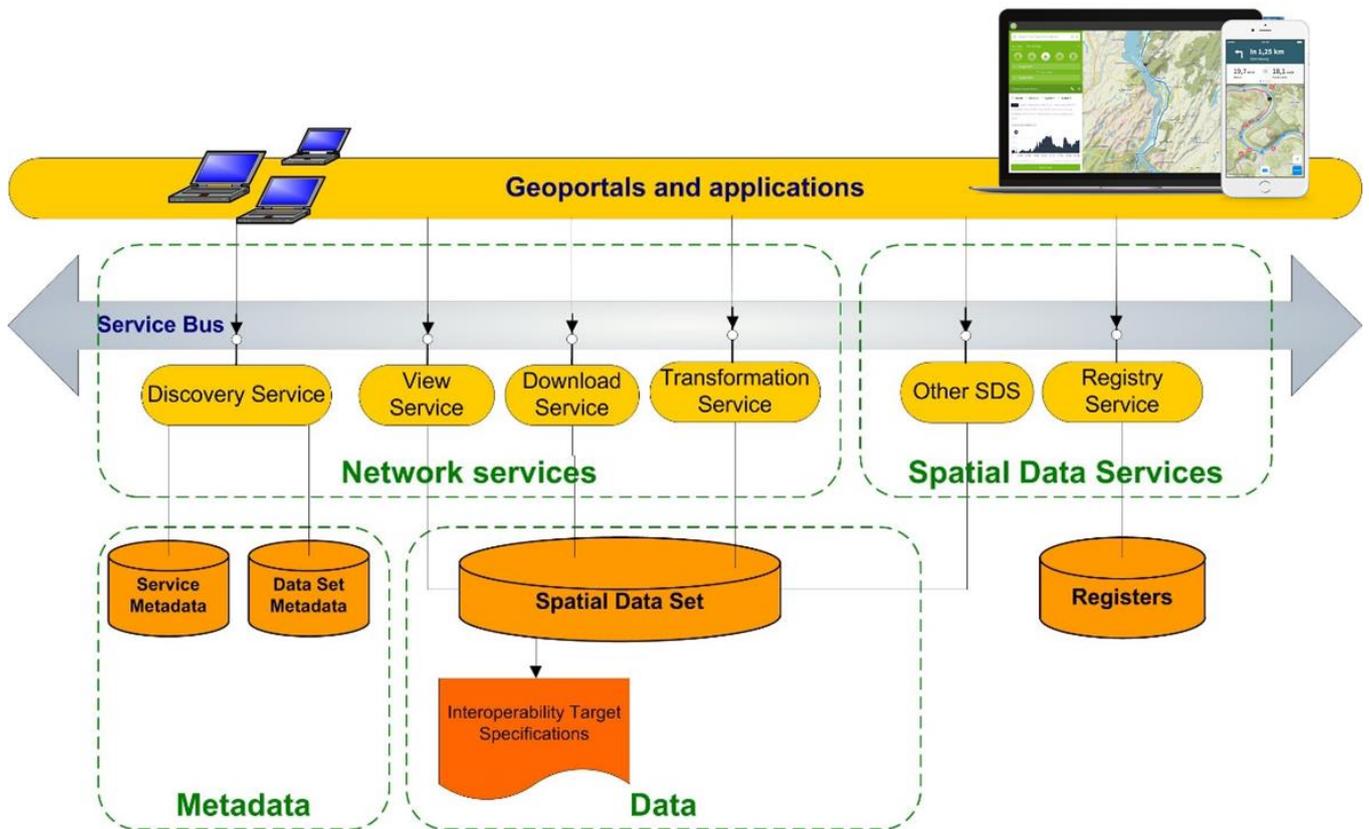
Figure 2. Distributed Service Oriented Architecture of INSPIRE. Source: European Commission, Joint Research Centre.

Deadlines for the implementation of the different components of the infrastructure are defined by the INSPIRE roadmap (see Figure 3), which foresees different milestones till 2021 according to the Annexes and the type of resources or services.

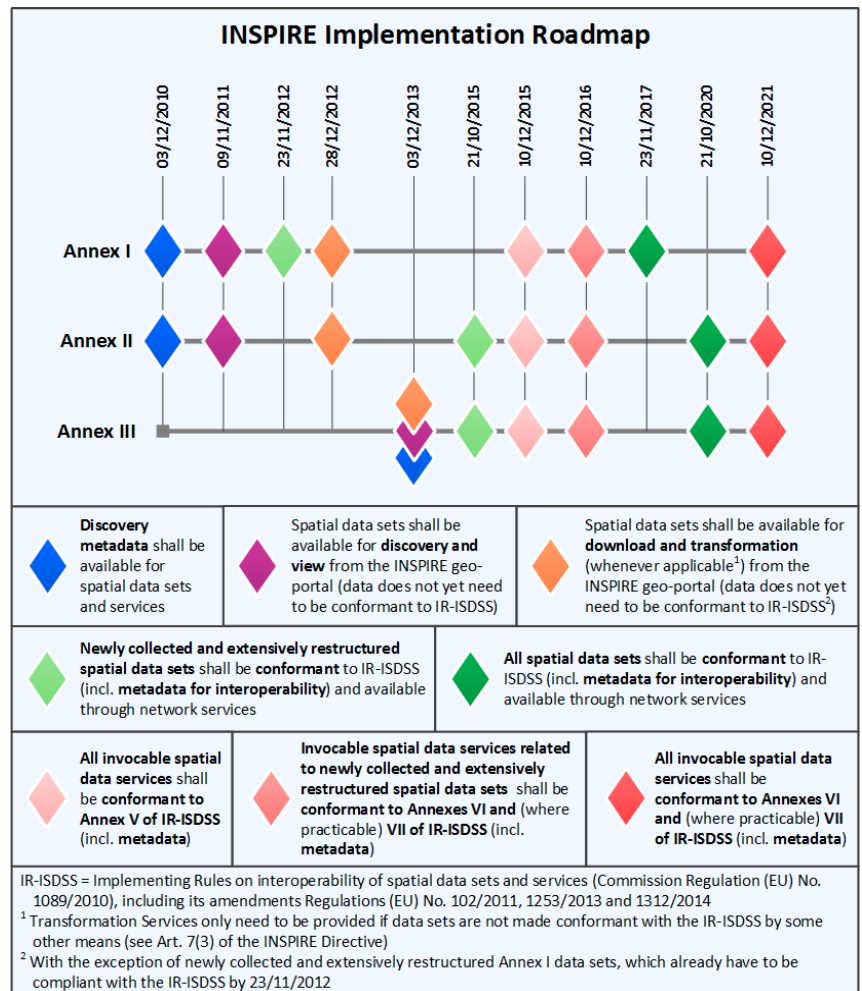

Figure 3. INSPIRE Implementation Roadmap. Source: European Commission, Joint Research Centre.

A number of important milestones have been already reached, however there are still activities to be completed, especially regarding data harmonisation and conformity which is crucial for the overall interoperability of the infrastructure. It goes in particular for Annexes II and III where the deadline for data harmonisation is set for the end of 2020. This means that, at the time of writing (beginning of 2020), data sets falling under related data themes are still made available mostly in a non-harmonised manner.

## 3.2 INSPIRE Geoportal

The entry point to the INSPIRE infrastructure is the INSPIRE Geoportal[20]. It serves as a central access point to the data and services from public organisations in the EU MS and EFTA countries which fall under the scope of INSPIRE. The INSPIRE Geoportal enables cross-border data discovery, access, visualisation and download. It does not store any geospatial data, but it simply acts as the main client application of the whole INSPIRE infrastructure by exposing data through the harvesting of the CSW endpoints made available by MS. Alongside the INSPIRE Geoportal, which is operated by the EC, there are also national geoportals operated by single countries. Links to national geoportals are available in the INSPIRE Knowledge Base (IKB) section entitled *INSPIRE in your country*[21].

---

[20] http://inspire-geoportal.ec.europa.eu
[21] https://inspire.ec.europa.eu/INSPIRE-in-your-Country

The first operational Geoportal Pilot was developed by the Joint Research Centre (JRC)[22] of the EC and released in 2011. In September 2018, a redesigned version was published (see Figures 4, 5 and 6) offering easier access to geospatial data in the EU. The new Geoportal was developed by the JRC in collaboration with and support from the EC Directorate-General for Environment[23], Eurostat[24] and the European Environment Agency[25]. It builds on the experience of running the Geoportal Pilot and supports several actions of the INSPIRE Maintenance and Implementation Work Programme[26], especially regarding improving the accessibility of data sets through Network Services and improving the availability of priority data sets for environmental reporting[27]. The redesigned Geoportal is a one-stop shop for public authorities, businesses and citizens to find, access and use geospatial data sets related to the environment in Europe. It also provides overviews of the availability of data sets by country and thematic area, and provides ready-to-use data either through interoperable web services or by direct download, to maximize their exploitation in third-party GIS clients and applications.

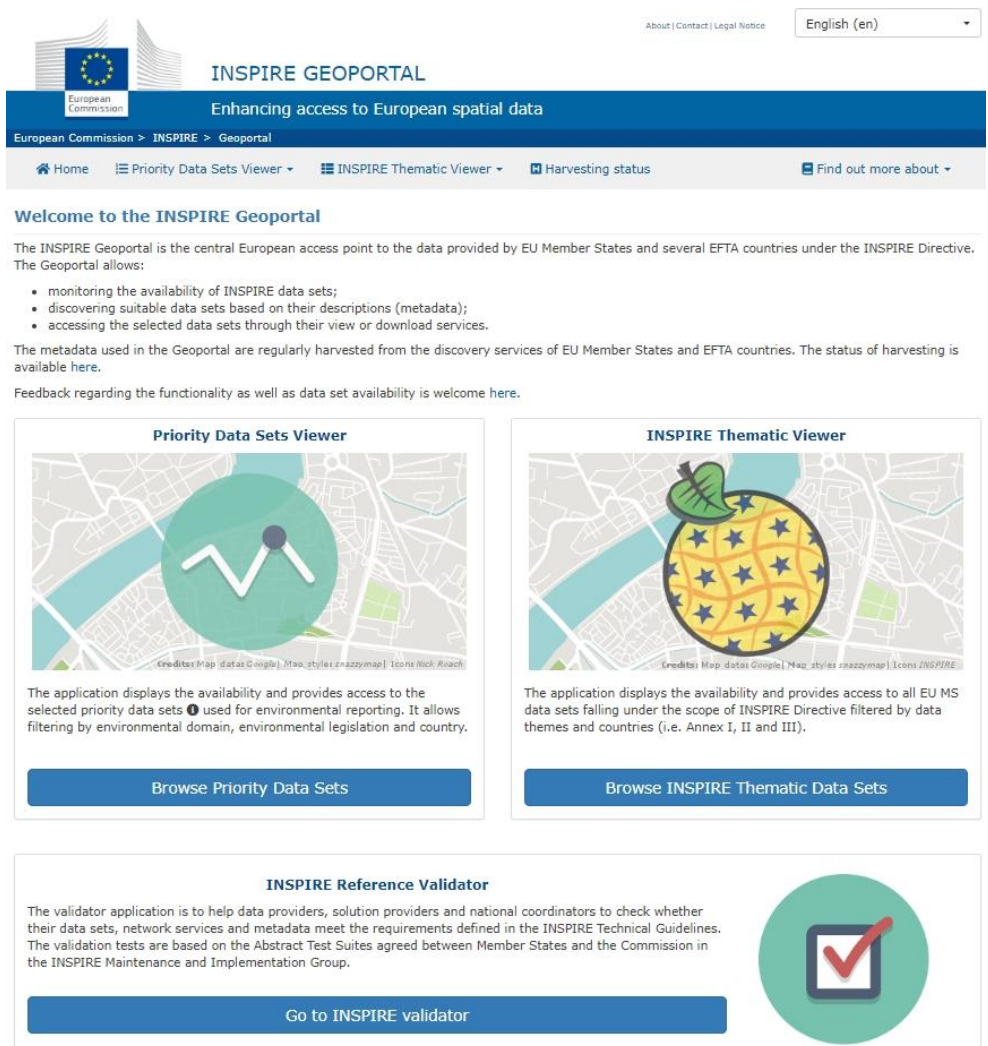

Figure 4. Landing page of the INSPIRE Geoportal. Source: European Commission, Joint Research Centre.

---

[22] https://ec.europa.eu/jrc/en
[23] https://ec.europa.eu/dgs/environment
[24] https://ec.europa.eu/eurostat
[25] https://www.eea.europa.eu
[26] https://webgate.ec.europa.eu/fpfis/wikis/pages/viewpage.action?pageId=268249090
[27] http://inspire.ec.europa.eu/metadata-codelist/PriorityDataset/

The Geoportal landing page provides access to 3 main applications:
1. *Priority Data Sets Viewer*, that displays the availability and provides access to the priority datasets used for environmental reporting[28];
2. *INSPIRE Thematic Viewer*, that displays the availability and provides access to all EU MS and EFTA countries datasets falling under the scope of the INSPIRE Directive, filtered by data themes and/or countries;
3. *INSPIRE Reference Validator*, a separate application that helps data providers check whether their data sets, services and metadata meet the INSPIRE requirements.

As mentioned above, the input source to the INSPIRE Geoportal is the harvesting of metadata from the officially registered Discovery Services of EU MS and EFTA countries. At the time of writing (January 2020) 37 Discovery Services are harvested on a regular basis (most of them weekly or monthly, however this is fully decided by the administrators of each Discovery Service), as shown in the *Harvesting status* section[29] of the Geoportal (see Figure 5).

Figure 5. Harvesting status on 15/12/2019, Source: European Commission, Joint Research Centre.

Insights into the current implementation status of the infrastructure are provided by the INSPIRE Thematic Viewer, which offers two possibilities for browsing datasets: by individual EU MS & EFTA country and by

---

[28] https://ies-svn.jrc.ec.europa.eu/projects/2016-5/wiki
[29] https://inspire-geoportal.ec.europa.eu/harvesting_status.html

INSPIRE data theme. Figure 6 shows the availability of datasets in EU MS and EFTA countries as of October 2019.

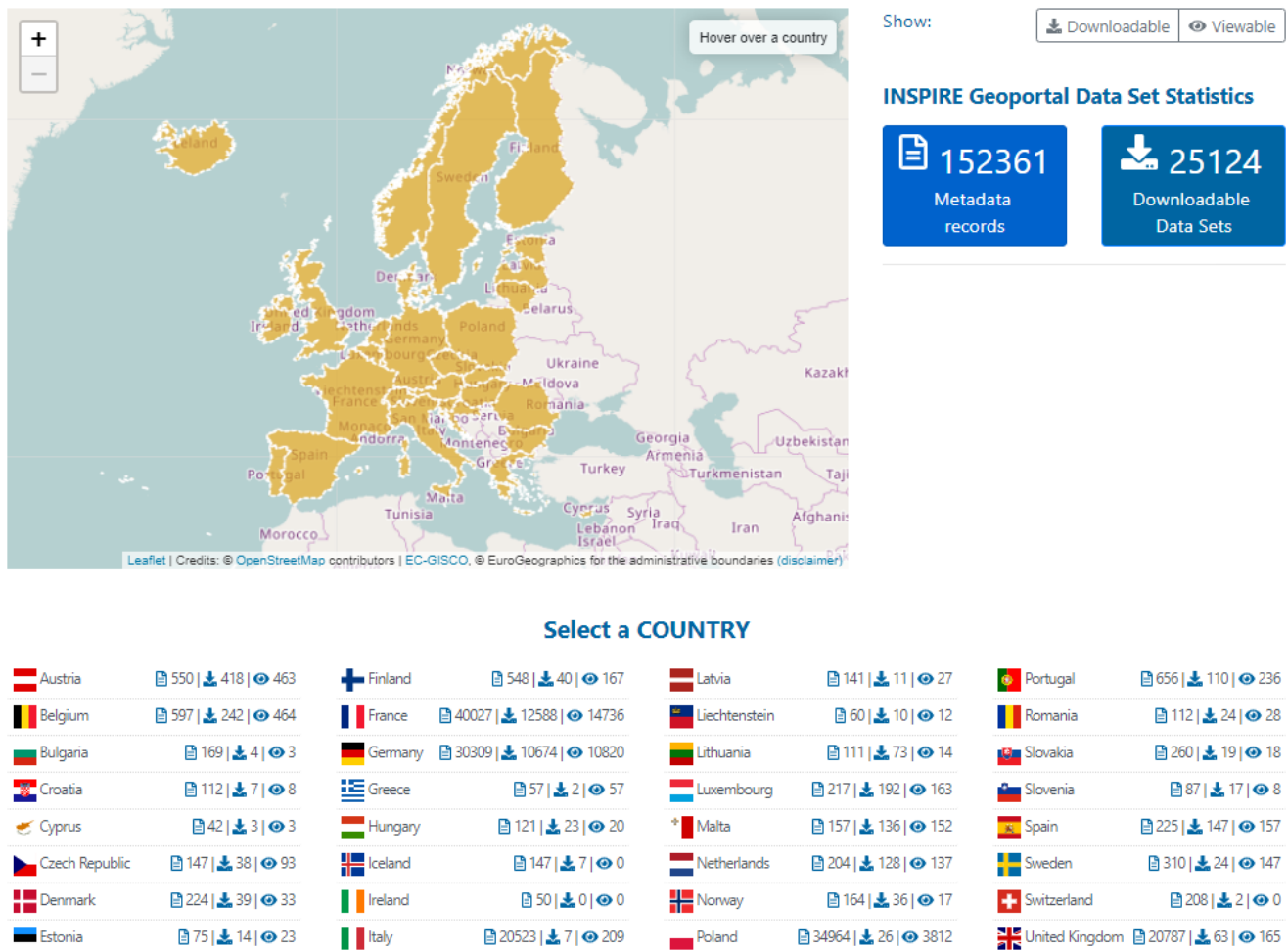

Figure 6. Availability of INSPIRE datasets in EU MS and EFTA countries in October 2019. Source: European Commission, Joint Research Centre.

The three numbers related to each country correspond to the number of available metadata records, downloadable datasets (i.e. data sets for which a Download Service is available) and viewable datasets (i.e. datasets for which a View Service is available). In the INSPIRE Geoportal about 150k datasets are available with metadata, of which about 24k are also viewable and about 13k downloadable (see Figure 6). The differences between the number of metadata records and the number of viewable and downloadable datasets demonstrate that the full implementation of INSPIRE is yet to be achieved. Similarly, for each EU MS and EFTA country the Priority Data Sets Viewer displays the availability of metadata, viewable and downloadable datasets for the priority datasets used for environmental reporting, which can be filtered by country, environmental domain or environmental legislation.

# 4. INSPIRE AS A BIG DATA INFRASTRUCTURE

## 4.1 Characteristics of INSPIRE in terms of Big Data

As already outlined, Big Data represents the information assets characterized by such a volume, velocity and variety to require specific technology and analytical methods for its transformation into value (De Mauro et al. 2015). In addition to volume, velocity, variety and value, characteristics of Big Data are veracity and visualisation, thus leading to the six Vs of Big Data. The Big Data landscape, including both the technologies (i.e. data lakes[30]) and infrastructures (i.e. databases and analytical tools) is evolving quickly. Geospatial data has always been considered to have more complex and larger datasets relevant for many other applications (McDougall and Koswatte 2018). The term Big Geospatial Data typically refers to spatial datasets exceeding the capacity of widespread computing systems. Many evidences have witnessed that a significant portion of Big Data is, in fact, Big Geospatial Data (Lee and Kang 2015). Spatial data comes from many sources and is used within many domains. According to the business model of the INSPIRE Directive, an efficient use of government resources requires that "spatial data are stored, made available and maintained at the most appropriate level" and that "it is possible to combine spatial data from different sources and share them between several users and applications" (European Parliament and Council 2007) – thus envisioning INSPIRE as a digital platform bringing together spatial data holders, analytics providers and users in sharing, combining and exploiting data. In Table 1, we map the six Vs of Big Data to the characteristics of data within the INSPIRE infrastructure.

Table 1. Characteristics of INSPIRE from the perspective of Big Data.

| Big Data Characteristics | INSPIRE Characteristics |
|---|---|
| Volume - refers to the size of data | The infrastructure includes geospatial data falling under 34 data themes from all EU MS, EFTA countries and some candidate and potential candidate countries. Size of data differs but many are large data sets covering up to the whole countries, commonly acquired and used in the form of raster imagery, point cloud data, sensor observations, etc. |
| Variety - refers to heterogeneous sources and the nature of data, both structured and unstructured | Geospatial data in the infrastructure are produced at different levels (municipalities, cities, regions and the whole countries). Some of them are harmonised and structured in an INSPIRE conformant way, but many of them are "as is" data sets. There are several different types of data, e.g. imagery data, geotagged text data, spatio-temporal observation data, structured and unstructured data, raster and vector data – many with complex structures. The INSPIRE *Find your scope* application[31] lists 338 spatial object types, 121 data types and 294 code lists/enumerations. |
| Velocity - refers to the speed of generation of data | The amount of geospatial data in the infrastructure is increasing on a daily basis as long as new resources are collected/produced and made available in MS Discovery Services. Data velocity can be monitored through the INSPIRE Geoportal. |

---

[30] https://en.wikipedia.org/wiki/Data_lake
[31] https://inspire-regadmin.jrc.ec.europa.eu/dataspecification/FindYourScope.action

| Veracity - refers to quality of data and data sources | The quality of geospatial data in the infrastructure varies from source to source and is expressed in relevant metadata elements e.g. Lineage. |
|---|---|
| Visualisation - refers to presentation of data of almost any type in a graphical format that makes it easy to understand and interpret them | While the INSPIRE Geoportal embeds a simple map-based visualisation, the datasets available in the infrastructure can be analysed and presented through any other visualisation tool suitable for geospatial datasets, including not only traditional ones such as charts, maps and imageries but also 3D models, animations, hotspot and change detection maps. The use of OGC interoperability standards to serve INSPIRE data potentially allows any client application to access and and visualise them. |
| Value - refers to the worth of the data being extracted or used | The extraction of information and the actual use of INSPIRE data has direct economic benefits for Europe. MS implementation reports clearly show such benefits (Cetl et al. 2017). |

Table 1 above shows that the INSPIRE infrastructure shares the characteristics of Big Data and thus it could be considered in all respects a Big Geospatial Data infrastructure. The INSPIRE geospatial interoperability principles offer a flagship model for integrating huge amounts of heterogeneous sources of public sector information originating from a variety of providers, domains, administrative levels and cultural borders.

## 4.2 Challenges from the user perspective

While several opportunities emerge from the establishment of the INSPIRE Big Geospatial Data infrastructure, at the same time there are also many challenges which especially pertain to the usability of such infrastructure. The most pressing ones are elaborated in more detail in the following subsections.

### 4.2.1 Discoverability of datasets

Fostered by INSPIRE, the development of NSDIs in Europe has made more and more geospatial resources (datasets and services) available on the web. The first visible component for users, which crucially enables them to search and retrieve resources is metadata (Cetl et al. 2016). Metadata records in INSPIRE are split into two, with individual records being created for (i) geospatial data and (ii) geospatial services, both served in the standardised way by using Discovery Services (usually OGC CSW). Users then search for the resources by using discovery clients, for example the INSPIRE Geoportal as well as popular GIS clients such as QGIS[32]. This works fine in an SDI environment where both data producers and users are aware of SDI principles, in particular the corresponding services, the specialised GIS tools and how to use them. However, many users (both mainstream ICT developers and end users) are not at all aware of SDIs. Those non-expert users typically search for geospatial resources through standard web search engines such as Google and Bing. In addition to that, there are still a lot of geospatial data producers who – instead of creating and publishing metadata in a predefined structure to make their resources discoverable through an SDI – simply make them available on the web without documentation and standardised publication. The solution to that could be the use of metasearch enhanced crawlers to collect online accessible geospatial resources published by OGC services. There are also experiences and good practices with landing pages of datasets and services that could be generated from catalogue metadata (rather than maintaining the information in different systems). Linked data also constitutes a fast emerging trend, with clear potential to benefit SDIs (Bucher et al. 2020), leveraging a way to interconnect related data resident on the web, and deliver it in a more effective manner to increase its value for users. The resulting "Web of data" has recently started being populated with geospatial data[33]. There are other ongoing discussions dedicated to web developers, spatial data publishers and search engine optimisation (SEO) experts related to search engines indexing optimisations when publishing geospatial data. A first, notable effort to create a search engine for geospatial resources is Google's Dataset Search[34] launched in 2018.

---

[32] https://qgis.org

### 4.2.2 Combining national datasets to create pan-European products

There is a growing demand for more and better quality data both within the EC and the EU MS to support a number of key policies related, but not limited to the environment. Often, several data sources are available at MS level representing the same spatial objects which differ in various characteristics and quality criteria such as cartographic scale, level of detail, positional accuracy, timeliness, update frequency and licensing conditions. The ultimate goal of INSPIRE is to have harmonised national datasets from MS that can be seamlessly used at cross-border and transnational levels and facilitate the creation of consistent pan-European datasets without or with limited additional processing. However, the latter is not among the original INSPIRE objectives. The latter should be provided based on performant and stable services in line with the INSPIRE requirements. The challenges in creation of such pan-European datasets are twofold, i.e. related to a technical and a non-technical harmonisation. In terms of technical harmonisation, reference data are not yet available for all MS from the INSPIRE Geoportal with harmonised physical data models and functioning services. Further challenges include harmonisation in terms of level of detail, scale and edge-matching; in addition, the scope of the datasets provided by MS in the INSPIRE infrastructure might differ a lot, ranging from single national datasets to multiple regional or even local datasets. From the non-technical perspective, despite the efforts made to harmonise the access conditions to spatial datasets and services (European Commission 2010; European Commission 2013), there is still a high diversity of licensing conditions between countries and in some cases even among national public authorities of the same country that overall might create serious legal obstacles to data access and reuse. Data licenses range from open data licenses such as those from Creative Commons[35] to more restrictive and in some cases custom or national-specific licenses. This diversity clearly hampers the joint use of datasets as well as the development of consistent and harmonised EU-wide products derived from national datasets. Even in the circumstance when multiple datasets from different MS or data providers are published under open data licenses, the diversity between these licenses might pose legal obstacles to their combination and joint use from third-party actors. In addition, sometimes MS organisations adopt different technical means to restrict access to data services (e.g. through authentication mechanisms or by imposing hefty access fees), thus further impacting on data accessibility and usability.

### 4.2.3 Data access and consumption by clients

When implementing INSPIRE, MS have adopted a number of different strategies for the implementation of network services. Often the heterogeneity of the European data landscape had led to a lack of agreement between data providers on how to organise Big Data for effective utilisation. Examples include the number of data sets grouped together within one (or few) View and Download Services as well as the criteria used to group such data sets (e.g. by data theme, geographic area, scale, use case or national provider/organisation). A first drawback of this approach is the difficulty for client applications to easily find the desired resources. In some cases Big Geospatial Data such as huge databases or coverages covering whole countries (e.g. national registries of addresses or national orthophotos), which typically correspond to files of gigantic size, are also served through one single service. The consequence on the client side is the overall difficulty in accessing such datasets served through Download Services such as WFS or Atom feeds, in particular the extremely long waiting times required for download. Even in the case that the download is successful, the ultimate consequence is still the issue of consuming (i.e. visualising, analysing and processing) such data for end users.

---

[33] http://ggim.un.org/meetings/GGIM-committee/8th-Session/documents/Standards_Guide_2018.pdf
[34] https://datasetsearch.research.google.com
[35] https://creativecommons.org

### 4.2.4 Cloud infrastructures

The components of an SDI can be integrated into the cloud as value-added services (Schäffer et al. 2010). Cloud infrastructures are now available as a cost-effective and efficient alternative to on-premises provision of INSPIRE services from providers such as Amazon and Microsoft (Bragg 2017). Only few MS have started to publish their INSPIRE resources in a cloud infrastructure and it can be expected that more data providers and organisations will adapt the same approach in the future, although cloud computing typically raises several concerns due to lack of trust and transparency. INSPIRE can ultimately benefit from the ability of cloud infrastructures to handle large amount of requests and deliver data in a robust and performing manner. By migrating services to the cloud, the geospatial resources provided by these services would be immediately available in a scalable fashion for on-demand use. The central components of the INSPIRE infrastructure, which are technically managed by the JRC, have also not yet been migrated to the cloud. The only exception is the above mentioned INSPIRE Reference Validator[36], which is deployed on the cloud since spring 2019 to address the increased user base while at the same time providing satisfactory performances.

## 5. CONCLUSIONS AND OUTLOOK

Since its adoption in 2007, the INSPIRE Directive has been the driver behind the development of an EU-wide Spatial Data Infrastructure based on the interoperability principles ultimately aiming at the creation of a single European (geo)data space. The existence of this SDI has been initially considered of primary importance in support of EU environmental policies and activities impacting the environment. However, location has become pervasive across multiple policy and societal domains and so is the relevance and potential of geospatial data. Accordingly, INSPIRE holds the potential to enable a full use of spatial information across the public sector, allow multiple stakeholders (including not only governmental agencies but also private companies, researchers and citizens) to access spatial data across Europe, assist cross-border policy-making and support better integrated public eGovernment services (Cetl et al. 2017). Thanks to the numbers, increasing on an almost daily basis, of metadata records, datasets and services shared by European countries, INSPIRE is gradually evolving into a reference European SDI whose data possess all the characteristics of Big Data (see Section 4). Given that INSPIRE as an SDI has been a pioneer of the European digital society and economy, and that many efforts to build SDIs even beyond Europe have looked at INSPIRE as a model, after more than ten years since INSPIRE inception a number of lessons from a Big Data perspective have been learnt. These include opportunities, threats and dependencies coming from the emerging technological, societal and economic trends, which should guide the future steps so that INSPIRE – subject to the necessary political support and mandate – can continue to play a key role within the European geospatial digital data revolution.

---

[36] http://inspire.ec.europa.eu/validator/about

From our perspective the notion of SDI as it was originally defined (Clinton 1994) is evolving. The processes which are typically described by SDIs are mainly linear, i.e. data is first collected (often only once) by specific actors (usually trained professionals), harmonised and published by governmental data providers and finally consumed by data users. Based on this, there has been the belief that once such an infrastructure was in place, people would simply use it. But, in the era of Big Data, reality has become much more complex. First, new actors such as private companies, citizens and researchers have become key players in terms of data collection, thus leading to the term 'produsers' to denote their blurred role of being both producers and users of data (Coleman et al. 2009). Nonetheless, the public sector remains a major actor, whose main advantage is the fact that large portions of its datasets are quality controlled and often rooted in the formalised data value chain processes that are legislatively defined. Second, data are currently collected at unprecedented speeds, and from sources such as drones, smartphones, in situ sensor networks and Internet of Things (IoT) devices which did not even exist when the term SDI was first coined. This increased amount of data brings obvious complexities when it comes to defining ownership, privacy, and licensing. In this new context, it is crucial that traditional SDIs evolve into modern data ecosystems. These can be defined as complex systems of people, organisations, technology, policies, and data in a specific area that interact with each other and their surrounding environment for a specific purpose. Such ecosystems evolve and adapt through a cycle of data creation and sharing, data analytics, and value creation in the form of new products, services, or knowledge, which, when used, produce new data feeding back into the ecosystem (Pollock 2011; UN Environment Assembly 2019; Oliveira et al. 2019). Thus, the key difference of data ecosystems compared to SDIs is the cyclical flow that links the processes of data creation and sharing, data analytics, value creation and use, in turn generating new data in a continuous feedback loop between the stakeholders involved. In other words, the processes within data ecosystem are mainly driven by specific use cases, address specific users and are described by dynamic and non-linear processes. From a Big Data perspective, data analytics combined with artificial intelligence are particularly important for improving policymaking and service delivery (Lisbon Council 2019).

We therefore anticipate the evolution of INSPIRE into a new data ecosystem for environment. In line with the recently-published priorities of the EC, which set out a European Green Deal for the EU and its citizens (European Commission 2019) as well as the strategy to establish a common European data space (European Commission 2018), INSPIRE should act as an integrated data ecosystem which, sitting on top of this horizontal EU data space and interconnected with the EOSC (i.e. with data and services from many different sources and actors, not necessarily geospatial), can deliver efficient solutions to ensure good policymaking in the environmental domain and beyond. Using another term to express the same idea, INSPIRE should evolve from a highly distributed and fragmented infrastructure into a centralized platform (Reitz 2019).

The transition towards a successful data ecosystem implies transitions in a number of dimensions. From the technological perspective, it is still crucial that such a data ecosystem is built on open standards. However, while a SOA based on traditional OGC web standards was at the forefront when INSPIRE was conceived, such standards no longer reflect the modern practices of data exchange through the web (Open Geospatial Consortium 2019). A new family of API-based standards, collectively called OGC APIs[37], is under development through a user-centric, data-driven approach to maximise the benefit for the future users. The first and currently only standard published, the OGC API - Features[38], has been already identified as a candidate standard on which a proposal for a specification for setting up INSPIRE Download Services is under development[39]. The OGC API - Features is a REST API that quickly and easily accesses geospatial features on the web, potentially allowing to overcome the issues described in Subsection 4.2.3. Being designed as a modern web standard, it is GeoJSON-oriented (although other encodings are also supported) and thus it goes in the same direction of defining alternative encodings to simplify and flatten the INSPIRE complex data models, an activity that the JRC has already started in 2018[40]. An increased flexibility, which – at least for selected spatial object types – relaxes some semantic requirements and only secures a basic level of interoperability, might definitely help improve the overall usability of the infrastructure. Similarly, the OGC SensorThings API standard[41] is proposed as an INSPIRE Download Service (Kotsev et al. 2018). SensorThings API is also based on REST principles and provides a simple yet powerful means for retrieval of observation data.

---
[37] http://www.ogcapi.org

In the same direction of simplifying data access and usability, which would address the issue described in Subsection 4.2.1, APIs are also identified by the Open Data Directive (European Parliament and Council 2019) as the required tool to publish high-value datasets. APIs have real potential to make the new generation of data ecosystems user-friendly and usable by developers and end-users, often not familiar with spatial web services, as building blocks to create third-party applications to generate additional value. In recent years, several European countries such as France, Germany, Sweden, Ireland and Croatia have developed an API-based approach as an integral part of their SDI or INSPIRE developments.

Non-technical transitions are also key for the success of INSPIRE as a data ecosystem. An open platform model characterised by cyclical data flows between the actors and stakeholders involved only works if it generates value for all of them. Thus, it will be increasingly important to move away from a traditional vision which only looks at the interface between data providers and users, and include other key stakeholders such as partners, software providers, companies selling value-added services or data analytics, and intermediary organisations that facilitate interactions within the platform. All of this without forgetting that the same stakeholders can play multiple roles. Finally, horizontal aspects such as data management, governance, protection and sharing issues will need to be prioritised. The latter includes addressing the INSPIRE licensing scheme which is currently a serious obstacle for a full exploitation of the infrastructure. Taking the Open Data Directive – which requires that high-value datasets are "made available for reuse with minimal legal restrictions and free of charge" – as a reference, a possible path towards an increased usability of the INSPIRE infrastructure could be to require the publication of specific datasets without any access obstacle (e.g. authentication or payment of a license fee) and under an open license that allows for re-use for any purpose, including commercial. In turn, this would facilitate the cross-border combination and the creation of pan-European products, at least for these specific datasets, thus solving the issue described in Subsection 4.2.2. Last but not least, the currently slow and only partial implementation of INSPIRE (already discussed in Section 3.2) should be addressed through a well-thought combination of regulatory interventions and other non legal measures, the latter including incentives, benefits and constructive competitions (Lisbon Council 2019).

---

[38] https://www.opengeospatial.org/standards/ogcapi-features
[39] https://github.com/INSPIRE-MIF/gp-ogc-api-features
[40] https://github.com/INSPIRE-MIF/2017.2
[41] https://www.opengeospatial.org/standards/sensorthings